# The Algorithms of Updating Sequential Patterns

Qingguo Zheng Ke Xu Shilong Ma Weifeng Lv
National Lab of Software Development Environment
Department of Computer Science and Engineer
Beijing University of Aeronautics and Astronautics, Beijing 100083
{zqg, kexu, slma, lwf}@nlsde.buaa.edu.cn

**Abstract**

Because the data being mined in the temporal database will evolve with time, many researchers have focused on the incremental mining of frequent sequences in temporal database. In this paper, we propose an algorithm called IUS, using the frequent and negative border sequences in the original database for incremental sequence mining. To deal with the case where some data need to be updated from the original database, we present an algorithm called DUS to maintain sequential patterns in the updated database. We also define the negative border sequence threshold: Min_nbd_supp to control the number of sequences in the negative border.

## 1. Introduction

Most research of data mining has focused on the problem of mining association rules [1,2,5,7,8]. Some research studies the time constraint data mining: sequence mining [3,4,5, 9]. In such a case, the data being mined has a timestamp, the data will increase with the time. If we re-run the algorithm of data mining to analyze the whole database including incremental data and original data, it is obviously inefficient and time consuming.

Cheung and Han et al. [10] first proposed an algorithm, called FUP (Fast Update), for the incremental mining association rules. Subsequently, other researchers have proposed many algorithms [11,14,15,16,17] to solve the incremental updating association patterns. But few researchers study the incremental updating problem of sequence mining. Several efficient algorithms for maintaining sequential patterns have been developed [12,13]. Nevertheless, the problem of maintaining sequential patterns is much more complicated than maintaining associations rules, because the sequence search is more complex than the item search and the number of candidates in sequence mining is more than the number of ones in association mining under the same conditions. In [13], an algorithm called ISM (Incremental Sequence Mining) was proposed based on SPADE approach [18], which can update the frequent sequences when new transactions and new customers are added to the database. It builds an increment sequence Lattice that consists of all the frequent sequences and the negative border sequences. When new data arrive, the incremental part is scanned once and the result of scanning the new data is merged into the Lattice. If the transaction database is very large, the size of negative border will be very large, which will consume a lot of memory [13]. In [12], an algorithm called ISE (Incremental Sequence Extraction) was proposed for mining frequent sequence, which generates candidates in the whole database by attaching the sequence of the incremental database to the frequent sequence of the original database. So it avoid keeping the large number of negative border sequences and re-computing those sequences when the data in the original database have been updated. However, since the ISE algorithm doesn't keep negative border sequences, it will need searching the database more. Furthermore, the ISE algorithm only extends the suffix of frequent sequence of the original database, but not extend the prefix of frequent sequence of the original database.

In this paper, we propose an efficient algorithm, called IUS (Incrementally updating sequences) for computing the frequent sequence when new data are added into the origin database. The IUS algorithm minimizes computing costs by reusing the negative border sequence and frequent sequence in the original database. In order to control the memory and time of the negative border sequence consuming, we define a new threshold for the negative border sequences, called the minimum support of negative border sequences, denoted by Min_nbd_supp. Only the support of sequences whose supports are less than Min_supp and greater than Min_nbd_supp can become a member of negative border sequence set. By adjusting the value of Min_nbd_supp, we can prune those negative border sequences whose values of support are small and that they have little effect on the computing results. Thus we can save the memory of the negative border sequences consuming. We also extend both the prefix and the suffix of the frequent sequences in the original database to generate the candidates in the updated database. The experiments on the alarm data of GSM show that the speedups about IUS algorithm and Robust_search [19] algorithm are between 1 and 9.

In most cases, the data that occurred a long time ago must be deleted from the database, so that the result of data mining can correctly reflect the current fact. Thus, we propose an algorithm, called DUS (Decreasingly Updating Sequences), to re-compute frequent sequences and negative border sequences in the updated database based on the result of previously mining the original database. Theorem 1 in Section 5 gives the necessary condition that any sequence $seq_m$ in the original database becomes a frequent sequence in the updated database. According to the Theorem 1, we can choose the best value for Min_nbd_supp, which is called Minimum frequent threshold, denoted by Min_freq

The rest of this paper is organized as follows. Section 2 we compare our work with the related works of the incremental mining. Section 3 formally defines the notion and definition in the paper. As to the incremental mining, we present the IUS (Incrementally Updating Sequences) algorithm in Section 4. On the other hand, we propose the DUS (Decreasingly Updating Sequences) algorithm for the decreasing mining in Section 5. We experimentally validate the IUS algorithm on the real GSM alarm data in Section 6. Finally, we conclude in Section 7.

## 2. Related Work

Thomas et al. [16] first proposed an algorithm, called ULI (Updatae Large Itemset), based on negative border set. However, the ULI algorithm didn't consider incrementally updating the sequential patterns. In this paper, we propose an algorithm called IUS (Incrementally Updating Sequences) for incrementally maintaining sequential patterns. The differences between IUS and ULI are described as follows. Firstly, the IUS algorithm considers the sequential patterns, while the ULI algorithm considers the association patterns. Secondly, the two algorithms use the negative border sequence for different purposes. According to the change of negative border set in the original database, the ULI algorithm decides whether to search the whole database. The IUS algorithm reduces the times of searching original database by keeping the negative border sequences. Furthermore, when the original database being deleted, some sequences in the negative border will become a frequent sequence in the updated database. Thus the DUS (Decreasingly Updating Sequences) use the negative border sequences as the new candidates in the database being deleted.

The differences between the IUS algorithm and the ISM (Incremental Sequence Mining) algorithm [13] are explained in the following. In order to control the volume of memory that negative border consume, the IUS algorithm defines the minimum threshold for negative border: Min_nbd_supp. Only if the support of the sequence in the database is greater than Min_nbd_supp, the sequence will become a sequence in the negative border. By adjusting the size of Min_nbd_supp, we can drop out some sequences to save the memory. But the ISM algorithm will keep all of negative border sequences, which mean that it can't control memory consuming of negative border sequences.

The differences between the IUS algorithm and the ISE (Incremental Sequence Extraction) algorithm [12] are explained below. First the IUS algorithm extends both the prefix and the suffix of the frequent sequence in the original database to generate new candidates in the updated database, while the ISE algorithm only consider extending the suffix of the frequent sequence in the original database. Second the IUS algorithm makes use of the negative border sequences, while the ISE algorithm doesn't.

## 3. The Notion and Definitions

**DB**: The original database which contains old time-related data.

**db**: The increment database which contains new time-related data.

**dd**: The decrement database from DB which contains deleted time-related data.

**U**: The updated database. When database being increasingly updated, the total set of data which are equal to DB+db .When database being decreasingly updated, the total set of data which are equal to DB-dd.

**Support(F, X):** the support of the sequence X in the X database, where $X \in \{db, dd, DB, U\}$.

**Min_supp** :Minimum support threshold of the frequent sequence.

**Min_nbd_supp**: Minimum support threshold of negative border sequence.

$C^X$: Candidate sequences in X database, where $X \in \{db, dd, DB, U\}$.

$L^X$ : Frequent sequences in the X database, where $X \in \{db, dd, DB, U\}$.

**NBD(X)**=$C^X$- $L^X$, where NBD(X) consists of the sequences in X database whose sub_sets are

Figure 1 Notion in the paper

**Definition 1. The negative border** is the collection of all sequences that are not frequent but all of whose sub-sequences are frequent [6]. The negative border in the database DB is noted by NBD(DB), and NBD(DB)=$\{\alpha \mid \alpha \in C^{DB}-L^{DB}$ and support($\alpha$)>Min_nbd_supp in the DB$\}$, where **Min_nbd_supp** is the threshold of the negative border in DB. Obviously Min_nbd_supp<Min_supp, thus the region of support (NBD(DB)) is [Min_nbd_supp, Min_supp].

**Definition 2.** Given an sequence $seq_m =< e_{i1}, e_{i2},\ldots,e_{im} >$ and a temporal database DB, the times of the sequence $seq_m$ occurring in the DB are defined as

occur ($seq_m$, DB)=| the times of $seq_m$ occurring in DB |

**Definition 3.** Given an sequence $seq_m =< e_{i1}, e_{i2},\ldots,e_{im} >$, The support of $seq_m$ in database DB is defined as

$$\text{support}\left(\text{seq}_m, \text{DB}\right) = \frac{\text{occur}\left(\text{seq}_m, \text{DB}\right)}{|\,\text{DB}\,|}$$

Obviously, we have

$$\text{occur}(\text{seq}_m, \text{DB}) = \text{support}(\text{seq}_m, \text{DB}) \times |\,\text{DB}\,|$$

**Property 1** Let B be a frequent sequence in DB, If $\forall A$, $A \subseteq B$, we have occur(A, DB)>occur(B,DB).

**Property 2** The lemma 5 proposed by Cheung and Han et al.[10], describe the necessary condition that the candidate set become a large itemsets in the updated database. In this paper, the lemma 5 is extended according to the time constrained, i.e., given a sequence S and $S \in L^{U}$(U=DB+db), then we have $S \in L^{DB}$ or $S \in L^{db}$.

**Proof:** we assume that $S \notin L^{DB}$ and $S \notin L^{db}$.

From the assumption, we have support(S, DB)≤occur(S, DB)•Min_supp,

support(S,db) ≤occur(S, db)•Min_supp,

Hence,support(S, DB)＋support(S,db) ≤(occur(S, DB)＋occur(S, db))•Min_supp i.e. support(S, DB+db) ≤ occur(S, DB+db) • Min_supp, which contradict the given condition: $S \in L^{U}$(U=DB+db). Therefore, $S \in L^{DB}$ or $S \in L^{db}$.

## 4. IUS (Incrementally Updating Sequences) algorithm

At first, we will introduce two lemmas as follows. Lemma 1 was first introduced in[12], which describe extending the suffix of the frequent sequence. Lemma 2 in this paper is proposed to describe extending the prefix of the frequent sequence in the original database. According to Lemma 1 and Lemma 2, we can generate new candidates for the updated database U by two times extending the sequences, i.e. $L^{DB} \times L^{db}$ and $L^{db} \times L^{DB}$. The definition and proof of Lemma 1 and Lemma 2 are described in detail as the following.

**Lemma 1**[12] Let F=<<D>,<$S_2$>>, where $S_2$ is a sequence in the incremental database db and D is a frequent sequence in the original database DB, and the sequence F is a frequent sequence in the updated database U. If $F \notin L^{DB}$, then the sequence $S_2$ occurs at least once in db.

**Proof:** 1.) If | F|=1, since $F \notin L^{DB}$, then F must occur in the db. Therefore, $S_2$ must occur in the db at least once.

2.) If | F|>1, then if $S_2$ doesn't occur in db, since $F \in L^{u}$, by the property 3, we know that $F \in L^{DB}$, which contradicts the given condition: $F \notin L^{DB}$. Therefore, the sequence $S_2$ occurs at least once in db.

**Lemma 2** Let F=<<$S_1$>,<D>>, where $S_1$ are a sequence in the incremental database: db and D is a frequent sequence in the original database: DB, and the sequence F is a frequent sequence in the updated database: U. If $F \notin L^{DB}$, then the sequence $S_1$ occurs at least once in db.

**Proof:** It can be proved as above.

We propose an algorithm, called IUS (Incrementally Updating Sequences), for incrementally updating sequential patterns. The IUS algorithm can be divided into two parts. The first part (From Line 1 to Line 40) is that the IUS algorithm uses the frequent and negative border sequences in DB and db as the candidates to compute new frequent sequences and negative border sequences in the updated database U. The second part (From Line 41 to Line 44) is that the IUS algorithm combine the frequent sequences in $L^{DB}$ which is frequent in the updated database U and not contained in $L^{db}$ with the frequent sequences in $L^{db}$ which is frequent in U and not contained in $L^{DB}$, i.e. $L^{DB} \times L^{db}$ and $L^{db} \times L^{DB}$, to generate the new candidates in the updated database U (Algorithm 1_1), and compute the supports of the new candidates in the updated database U (Algorithm 1_2). Algorithm 1_2 is the Robust_search Algorithm described in [19].

The Robust_search Algorithm does compute the support of sequences in the alarm event queue [19], especially, which can search the support of sequences from alarm event queue containing noise data. But because we mainly study the problem of updating sequential patterns, then we only consider the condition that the alarm queue doesn't contain noise data in this paper.

Algorithm 1_1 combines the sequences in $L^{DB}$ with the sequences in $L^{db}$ to obtain the new negative border NDB(U) in the updated database U. According to Property 2, we know that the IUS algorithm (from Line 1 to Line 40 in IUS) will have got the frequent sequences in the updated database U, by adopting the sequences in $L^{DB}$ and $L^{db}$ as the candidates for the updated database U. Therefore, we mainly get the new negative border sequences in the updated database U by the algorithm 1_1. According to Lemma 1 and Lemma 2, we know that the IUS algorithm can generate new candidates by exchanging the two sequences order, i.e. $L^{DB} \times L^{db}$ and $L^{db} \times L^{DB}$ (from Line 1 to Line 17).

We also define the threshold of negative border: Min_nbd_supp. By adjusting the value of Min_nbd_supp, we can control the number of negative border sequences to keep in memory. Therefore, we can save the computer resource by correctly setting the value for Min_nbd_supp. The following experiments in Section 5 show that the Min_nbd_supp can also improve the performance of the IUS algorithm (in Figure 3).

**Example 1.**

This example will illustrate the process of the IUS algorithm. The original database is DB, the incremental database is db, and the updated database is U.

1. DB

| Item | a | b | c | d | e | f | g | h | k |
|---|---|---|---|---|---|---|---|---|---|
| count | 4 | 3 | 2 | 2 | 1 | 1 | 1 | 1 | 1 |

Min_count=2, which is the minimum count of sequence becoming a frequent sequence in DB.

$L_1^{DB}$={<a,4>,<b,3>,<c,2>,<d,2>} $NBD_1$(DB)={<e,1>,<f,1>,<g,1>,<h,1>,<k,1>}
$L_2^{DB}$={<ab,2>,<bc,2>,<cd,2>} $NBD_2$(DB)={<ac,1>,<ad,1>,<bd,1>}
$L_3^{DB}$={<abc,2>} $NBD_3$(DB)={<bcd,1>}

2. db

| Item | a | f | b | d | c | e | g | h | k |
|---|---|---|---|---|---|---|---|---|---|
| count | 3 | 3 | 2 | 1 | 1 | 1 | 1 | 1 | 1 |

Min_count=2, which is the minimum count of sequence becoming a frequent sequence in db.

$L_1^{db}$={<a,3>,<b,2>,<f,3>} $NBD_1$(db)={<c,1>,<d,1>,<e,1>,<g,1>,<h,1>,<k,1>}
$L_2^{db}$={<ab,2>,<bf,2>} $NBD_2$(db)={ <bf,1>,<af,1>}
$L_3^{db}$={<abf,2>}

3.U=DB+db

| Item | a | b | d | f | c | e | g | h | k |
|---|---|---|---|---|---|---|---|---|---|
| count | 7 | 5 | 4 | 4 | 3 | 2 | 2 | 2 | 2 |

Min_count=4, which is the minimum count of sequence becoming a frequent sequence in U.

$L_1^{U}$={<a,7>,<b,5>,<d,4>,<f,4>} $NBD_1$(U)={<c,3>,<e,2>,<g,2>,<h,2>,<k,2>}
$L_2^{U}$={<ab,4>,<bd,4>} $NBD_2$(U)={<bd,3>,<ad,2>,<da,1>,<df,2>,<fd,1>}
$NBD_3$(U)={ <abd,3>}

**Algorithm 1 . IUS(Incrementally Updating Sequences)**

Input: Transaction database DB, db, U=DB+db, Min_supp, Min_nbd_supp
Output: Frequent alarm sequences set: $L^U$, Negative border sequences set :NBD(U).

```
Generate L^U_1 from L_1^DB , L_1^db, NBD(DB),NBD(db);
m=2; L_Size=0;
while ((|L^U| - L_Size)>0)
Begin
L_Size=| L^U |
For all seq_m ∈ L^DB and all subsets of seq_m are frequent in U
{
 If (seq_m ∈ L^db) Get occur(seq_m , db) from L^db ;
     Else If (seq_m ∈ NBD(db)) Get occur(seq_m , db) from NBD(db);
       Else search occur(seq_m , db) in db ; /* compute frequent sequence in L^DB */
If ((occur(seq_m , db)+ occur(seq_m , DB))>(Min_supp• |U|))
                 Insert seq_m into L^U;
     else { Prune the seq_m and the sequence containing seq_m from L^DB and NBD(DB);
             If((occur(seq_m , db)+ occur(seq_m , DB))>(Min_nbd_supp• |U|))
                                      Insert seq_m into NBD(U);
          }
 }
For all seq_m ∈ L^db and seq_m ∉ L^DB and all subsets of seq_m are frequent in U
 {
  If(seq_m ∈ NBD(DB)) Get occur(seq_m , DB) from NBD(DB)
      Else search occur(seq_m , DB) in DB ;/* compute frequent sequence in L^db */
  If( (occur(seq_m , db)+ occur(seq_m , DB))>(Min_supp• |U|) )
                               Insert seq_m into L^U ;
      else { Prune the seq_m and the sequence containing seq_m from L^db and NBD(db);
                 If((occur(seq_m , db)+ occur(seq_m , DB))>(Min_nbd_supp• |U|))
                                            Insert seq_m into NBD(U);
           }
 }
 For all seq_m ∈ NBD(DB) and seq_m ∉ L^db and seq_m ∉ NBD(db)
    and all subset of seq_m are frequent
 {   Search occur(seq_m , db) in db
     If((occur(seq_m , db)+ occur(seq_m , DB))>(Min_nbd_supp• |U|))
                                  insert seq_m into NBD(U);
 }
For all seq_m ∈ NBD(db) and seq_m ∉ L^DB and seq_m ∉ NBD(DB)
   and all subset of seq_m are frequent
   {    Search occur(seq_m ,DB) in DB
        If((occur(seq_m , db)+ occur(seq_m , DB))>(Min_nbd_supp•|U|))
                                 insert seq_m into NBD(U);
   }
 / * generate new negative border from L^db and L^DB */
  generate Negative Border NBD_m( U ). from L^U_m-1;  /* Algorithm 1_1 */
m=m+1;
end. /* end of while */
```

**Algorithm 1_1**

Input: Frequent sequence $L^{U}_{m-1}$ , $L^{U}_{m}$

Output: Negative Border $NBD_m(U)$.

$C_m=\Phi$;

For all $\alpha_{m-1} \in L^{U}_{m-1}, \beta_{m-1} \in L^{U}_{m-1}$, where $\alpha_{m-1} \in L^{DB}_{m-1}$ , $\beta_{m-1} \in L^{db}_{m-1}$ , $\alpha_{m-1} \neq \beta_{m-1}$ ,and

$\alpha = < e_{i1}, e_{i2}, \ldots, e_{im-1} >$ , $\beta = < e_{i1}`, e_{i2}`, \ldots, e_{im-1}` >$

{

/* generate candidate $C_m=\alpha_{m-1} \cap \beta_{m-1}$ */

If ($e_{i2}= e_{i1}` \cap e_{i3}= e_{i2}` \cap \ldots \cap e_{im-1}= e_{i\,m-2}`$ ) then

{

generate $\gamma = < e_{i1}, e_{i2}, \ldots, e_{im-1}, e_{im-1}` >$; /* Candidate generate */

$C_m= \{ \gamma \mid$ For all $L \subseteq \gamma$ and $|L|=m-1$ ,where $L \in L^{U}_{m-1} \}$; /*Pruning Candidate */

}

/* generate candidate $C_m=\beta_{m-1} \cap \alpha_{m-1}$ */

If ($e_{i2}`= e_{i1} \cap e_{i3}`= e_{i2} \cap \ldots \cap e_{im-1}`= e_{i\,m-2}$ ) then

{

generate $\gamma = < e_{i1}`, e_{i2}, \ldots, e_{i\,m-1} >$; /* Candidate generate */

$C_m= \{\gamma \mid$ For all $L \subseteq \gamma$ , $|L|=m-1$, where $L \in L^{U}_{m-1} \}$; /*Pruning Candidate */

}

} /* end of for loop */

For all $\alpha \in C_m$

{ if($\alpha \notin L_m^{U}$) /* U=DB+db */

{ Search occur($\alpha$, U) in U; /*Algorithm 1_2*/

If(occur($\alpha$ , db)>(Min_nbd_supp*|U|)) Insert $\alpha$ into NBD(U) ;

}

}

## 5. DUS (Decreasingly Updating sequences) algorithm

In most situations, the data that occurred a long time ago will have side effects on the result of sequence mining. In order to guarantee that the result of data mining can really reflect the fact in time, some old data need to be deleted from the original database. So we propose an algorithm, called DUS (Decreasingly Updating Sequences), to deal with the problem. According to Theorem 1 that will be introduced below, the DUS algorithm select the sequence from the negative border sequences and frequent sequences in the original database as the candidates in the updated database to get the frequent sequences and negative border sequences in the updated database.

In the following, Theorem 1 gives the necessary condition that any sequence in the original database becomes a frequent sequence in the updated database

**Theorem 1.** Let DB denote the original database, dd denote the updated database, then the updated database U=DB-dd. Given that $seq_m$ is any sequence in the original database DB, the necessary condition that the sequence $seq_m$ is a frequent sequence in database U is that

$$\text{support}\left(seq_m, DB\right) \geq \text{Min_freq} ,$$

$$\text{where } \text{Min_freq} = \text{Min_supp} \times \frac{|DB| - |dd|}{|DB|}.$$

**Proof:**

Let D=occur($seq_m$ , DB), d=occur($seq_m$, dd), then occur($seq_m$, U)=D-d.

If the sequence $seq_m$ is a frequent sequence in the database DB, then we have

$$\text{support}(seq_m, U) = \frac{D - d}{|DB| - |dd|} \geq \text{Min_supp}$$

From the formula above and the definition of support($seq_m$ ,DB), we have

$$\text{support}(seq_m, DB) = \frac{D}{|DB|} \geq \text{Min_supp} \times \frac{D}{D - d} \times \frac{|DB| - |dd|}{|DB|}.$$

From the definition, d≥0, thus $\frac{D}{D - d} \geq 1$ .Therefore, we have

$$\text{support}(seq_m, DB) \geq \text{Min_supp} \times \frac{|DB| - |dd|}{|DB|}, \text{ i.e. } \text{support}(seq_m, DB) \geq \text{Min_freq}.$$

This completes the proof.

By Theorem 1, we know that only these sequences, whose supports in the original database are greater than minimum frequent threshold:Min_freq, could become a frequent sequence in the updated database U. So if Min_nbd_supp≤ Min_freq, the DUS algorithm only uses the sequences in the negative border and frequent sequences whose support are greater than Min_freq as the candidates in the U. If Min_nbd_supp> Min_freq, the DUS algorithm directly uses the sequences in n the negative border sequences and frequent sequences as the candidates in the database U, to compute new negative border sequences and frequent sequences in the updated database U.

In summary, the DUS (Decreasingly Updating Sequences) algorithm is divided into two cases below.

- If Min_nbd_supp ≤ Min_freq, $\forall seq_m \in NBD(DB) \cup L^{DB}$ and support($seq_m$,DB) ≥ Min_freq, the algorithm compute support($seq_m$,U) to decide that the sequence $seq_m$ belong to NBD(U) or $L^U$(U=DB-dd).
- If Min_nbd_supp> Min_freq, $\forall seq_m \in NBD(DB) \cup L^{DB}$, the algorithm compute support($seq_m$,U) to decide that the sequence $seq_m$ belong to NBD(U) or $L^U$(U=DB-dd).

In the application, we can estimate the distribution of the minimum frequent threshold: Min_freq. When we choose the value for Min_nbd_supp, we had better select the value that is equal to Min_freq . If Min_nbd_supp is less than Min_freq, by Theorem 1, we know that the negative border NBD(U) may contain many sequences that couldn't become a frequent sequence in the updated database. On the other hand, if Min_nbd_supp is greater than Min_freq, by Theorem 1, some sequences will be dropped out in the NBD(U), although some of them will become a frequent sequences in the updated database. In fact, Theorem 1 give the best value for Min_nbd_supp.

**Alogrithm 2 DUS (Decreasingly Updating Sequences)**

Input: Database dd, DB, U(U=DB-dd), $L^{DB}$, NBD(DB)

Output: $L^U$, NBD(U)

```
1.  For all seq_m ∈ L^DB
2.  begin
3.  Search  occur(seq_m , dd);  /* compute frequent sequence in L^DB*/
4.        If (occur(seq_m , DB)-occur(seq_m , dd))>(Min_supp* (|U|-|dd|)))
5.                      then insert seq_m  into  L^U
6.            else  If (occur(seq_m , DB)-occur(seq_m , dd))>(Min_nbd_supp* (|U|-|dd|)))
7.                      then insert seq_m  into  NBD(U);
8.  end
```

```
For all seq_m ∈ NBD(DB)
  begin
    Search seq_m  in  dd
    If(ocur (seq_m,DB)-occur (seq_m,dd) >(Min_supp* (|U|-|dd|)))
                    then insert seq_m into  L^U
            else if (ocur (seq_m,DB)-occur (seq_m,dd) >(Min_supp* (|U|-|dd|)))
                    then insert seq_m into into NBD(U);
   end
If(Min_nbd_supp≤Min_supp•(|DB|-|dd|)/|DB|  then
  For all seq_m∈NBD(DB)  and  occur(seq_m,DB)≥ Min_supp•(|DB|-|dd|)
  Begin
      Search seq_m in  database dd;
     If(occur (seq_m,DB)-occur (seq_m,dd) >(Min_supp* (|U|-|dd|)))
                    then insert seq_m into  L^U
            else if (ocur (seq_m,DB)-occur (seq_m,dd) >(Min_supp* (|U|-|dd|)))
                    then insert seq_m into into NBD(U);
  End
 If(Min_nbd_supp>Min_supp•(|DB|-|dd|)/|DB|)  then
  For all seq_m∈NBD(DB) and support(seq_m ,DB) ≥ Min_supp • (|DB|-|dd|)/|DB|
     Begin
      Search seq_m in  database dd;
      If(occur (seq_m,DB)-occur (seq_m,dd) >(Min_supp* (|U|-|dd|)))
                    then insert seq_m into  L^U
            else if (ocur (seq_m,DB)-occur (seq_m,dd) >(Min_supp* (|U|-|dd|)))
                    then insert seq_m into into NBD(U);
     End
```

## 6. Experiments

We conducted a set of experiments to test the performance of IUS(Incrementally update sequences) algorithm. The experiments were on the DELL PC Server with 2 CPU Pentium II,CPU MHz 397.952211, Memory 512M, SCSI Disk 16G. The Operating system on the server is Red Hat Linux version 6.0.

The data in experiments are the alarms in GSM Networks, which contain 181 alarm types and 90k alarm events. The time of alarm events range from 2001-03-15-00 to 2001-03-19-23. According the time of alarm occurring, we sort the alarm events in alarm queue [19]. Then we divide alarm events into 20k, 30k, 40k, 50k, 60k, 70k, 80k, 90k with the increment of 10k alarm events i.e. |dd|=10k. In the Figure 2, the broken line graph is denoted by nbdX, where X is the Min_nbd_supp. In the Figure 3, the broken line graph is denoted by suppX, where X is the Min_supp.

We compare the execution time of the IUS algorithm with that of the Robust_search algorithm in [19] on the whole database DB+dd. The speedup ratio is defined as: speedup=the execution time of Robust_search / the execution time of IUS.

In Figure 2, with the increment of the number of alarms, the value of speedup will increase. The speedup will decrease after reaching the largest value. The execution time of the IUS algorithm mainly depends on the number of different sequence between $L^{DB}$ and $L^{db}$, If the number of different sequences between $L^{DB}$ and $L^{db}$ is large, the IUS algorithm will consume more time. In Figure 2, with the increment of value of Min_supp, the speedup will increase. Because with the increment of value of Min_supp, the number of frequent sequences will decrease, then the number of different frequent sequences between DB and db will decrease too.

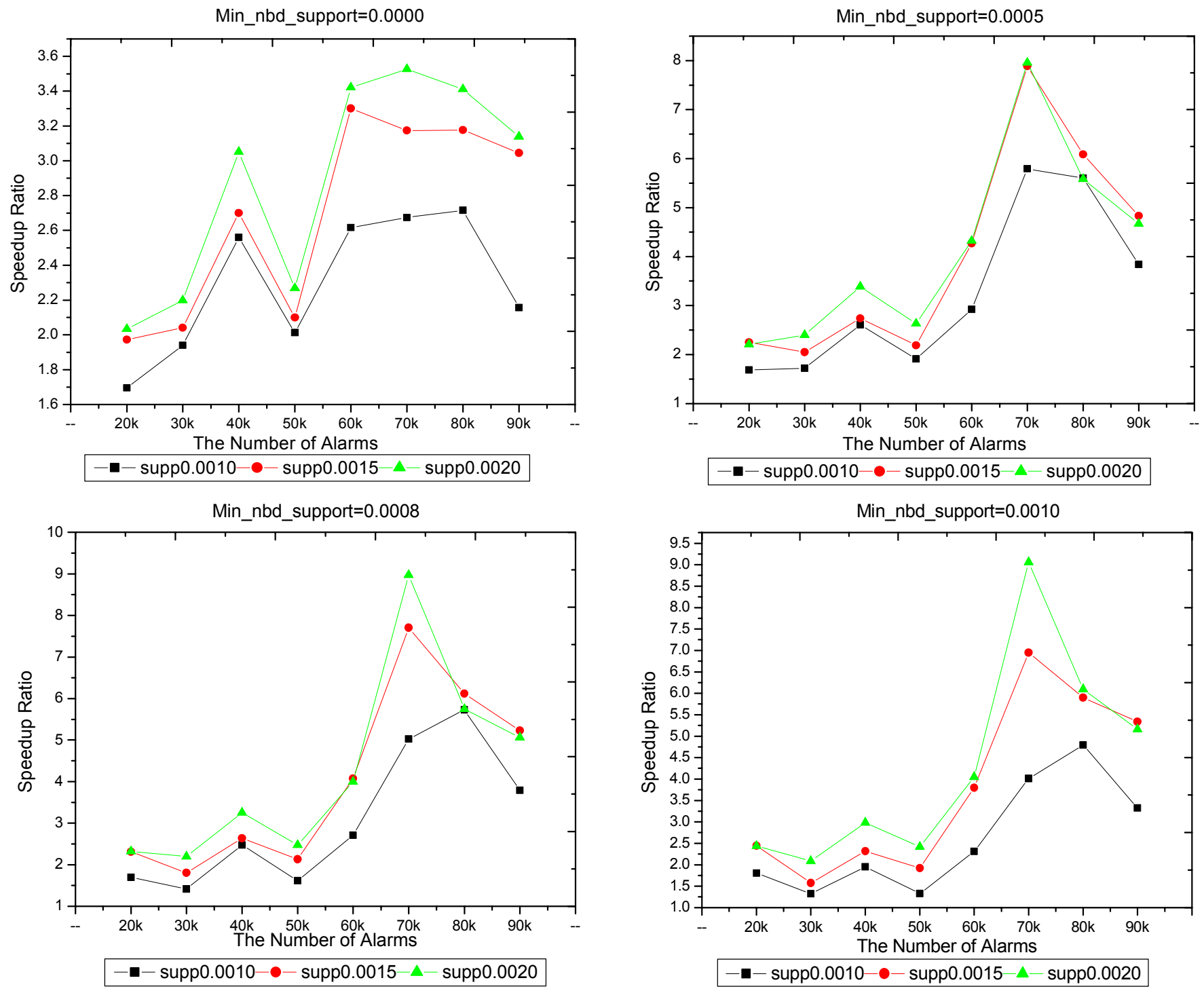

Figure 2 The Speedup Ratio at the Min_nbd_supp threshold

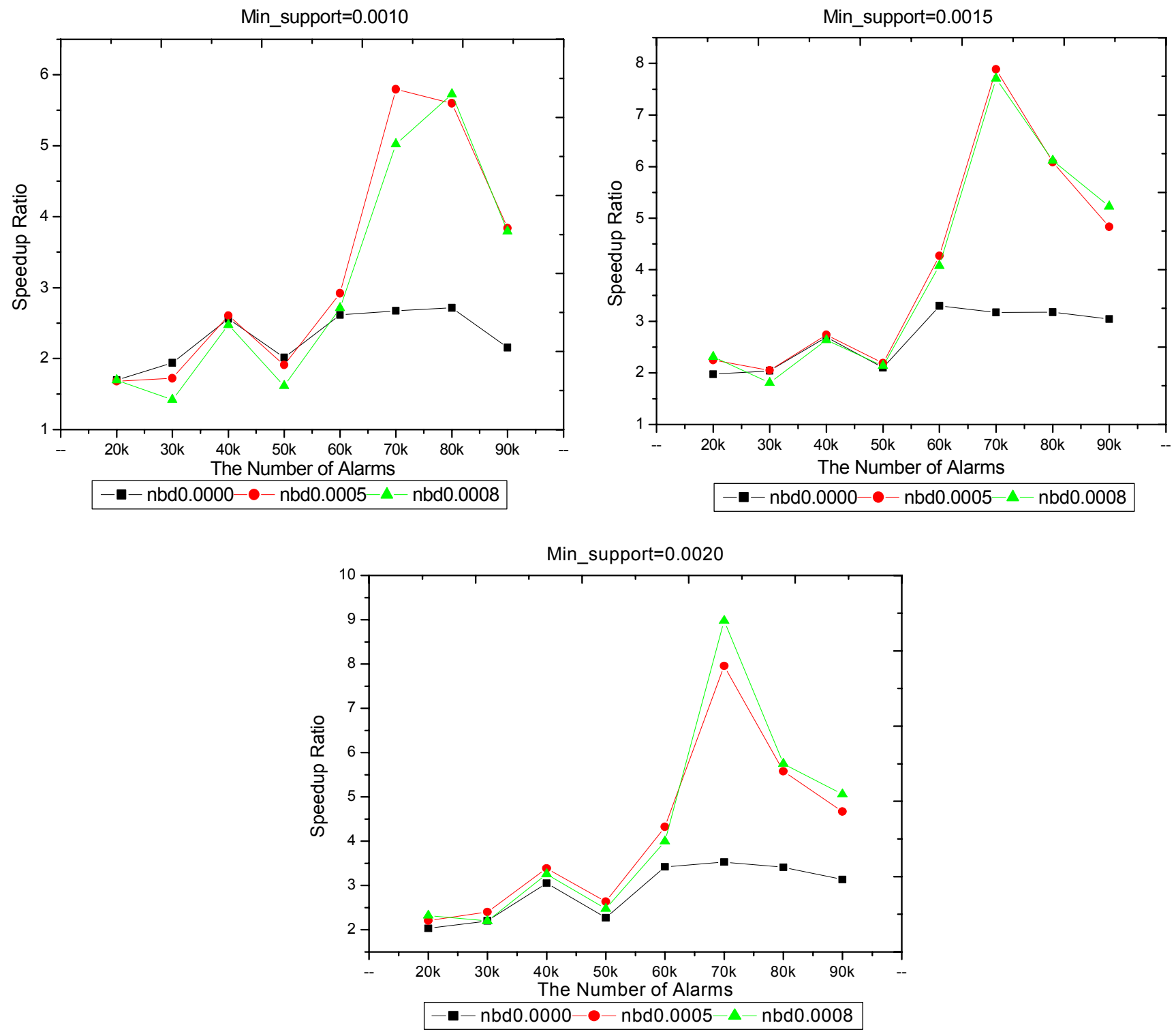

Figure 3 The Speedup Ratio at the Min_supp

In Figure 3, as the value of Min_nbd_supp increases, the speedup will also increase. As the value of Min_nbd_supp increase, the number of negative border sequences will decrease, then the number of negative border sequence between DB and db will decrease. Thus the IUS will consume less time and the speedup will increase.

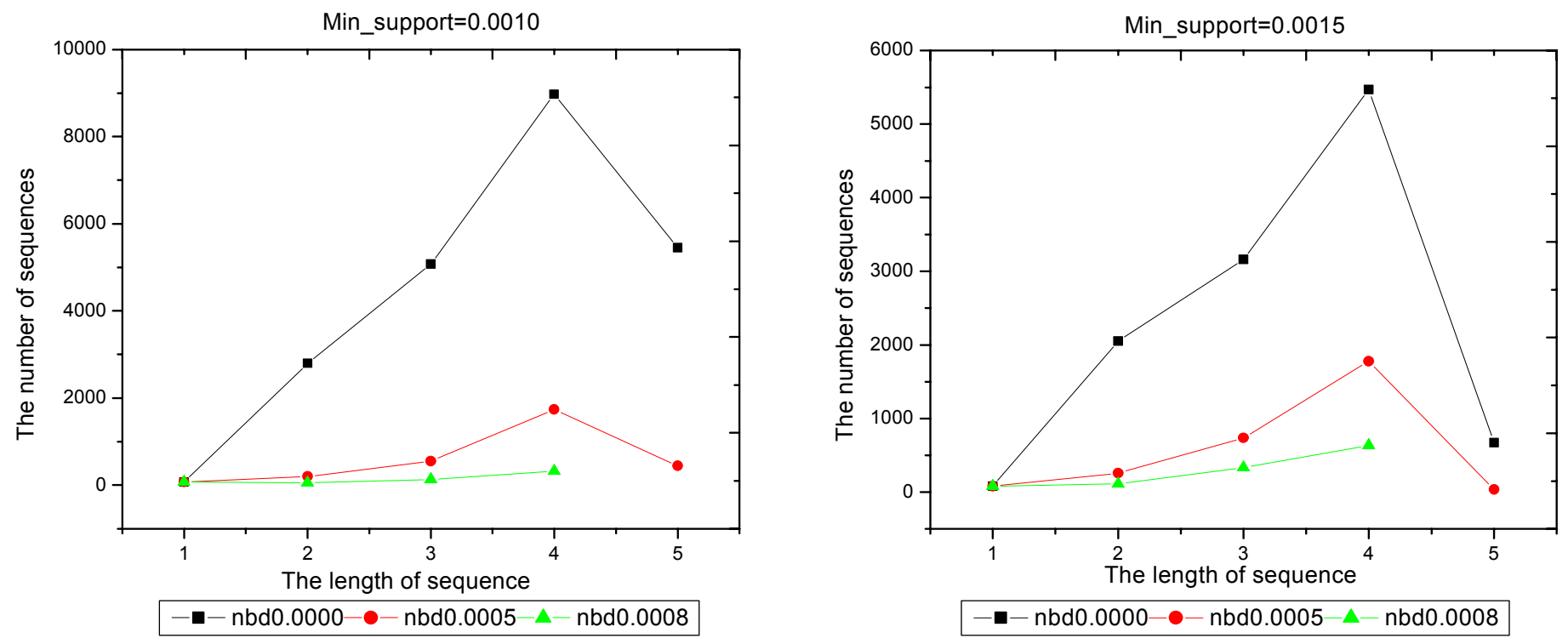

Figure 4 The number of negative border with different Min_nbd_supp

In figure 4, the data of the experiment is 20k of alarm events. With the increment of Min_nbd_supp, the number of negative border sequences is greatly reduced at the same value of Min_support. So we save the resources of computer by properly adjusting the value of Min_nbd_supp.

## 7. Conclusion

In this paper we present the IUS approach for incremental mining of sequential patterns in a large database. The results of experiments show that the IUS algorithm performance better than re-run the Robust_ search algorithm [19]. When to update the sequential pattern is another interesting subject in the future .

**ACKNOWLEDGEMENTS**

This research was supported by **National 973 Project of China Grant No.G1999032701 and No.G1999032709**. Thanks Professor Wei Li for the choice of subject and guidance of methodology. Thanks for the suggestions from Professor YueFei Sui of Chinese Academy Sciences. The author would like to thank the members of National Lab of Software Development Environment.